# Projectile Fragmentation of relativistic nuclei in peripheral Collisions

Swarnapratim Bhattacharyya[1], Maria Haiduc[2], Alina Tania Neagu[2] and Elena Firu[2]

[1]Department of Physics, New Alipore College, L Block, New Alipore, Kolkata 700053, India

Email: swarna_pratim@yahoo.com

[2]Institute of Space Science, Bucharest, Romania

## Abstract

A study of multiplicity distribution of singly charged, doubly charged, multi charged projectile fragments and shower particles have been carried out for the peripheral events of $^{16}$O-emulsion,$^{22}$Ne-emulsion and $^{28}$Si-emulsion interactions at an incident momentum of $(4.1-4.5)$ AGeV/c. Events having no target fragments have been designated as peripheral events. Percentage of peripheral events (POE) has been found to increase with the increase of projectile mass $A_P$ according to the relation $POE = (6.74 \pm 0.06)A_P^{(0.198 \pm 0.003)}$.  Our study reveals that the average multiplicity of shower particles ( $\langle N_s \rangle$ )  in peripheral events can be expressed as a function of the projectile mass $A_P$ according to the relation $\langle N_s \rangle = (1.23 \pm 0.09)A_P^{(0.24 \pm 0.07)}$.  On the other hand in peripheral events average multiplicity of multi charged fragments ($\langle N_F \rangle$) varies with the mass number of projectile beam according to the relation $\langle N_F \rangle = (0.065 \pm 0.005)A_P^{(0.76 \pm 0.07)}$.  The dispersion of multiplicity distribution for singly charged, doubly charged projectile fragments and also for the shower particles increase with the increase of projectile mass. Dispersion values of the multiplicity distribution for the multi-charged fragments decreases with the increase of projectile mass. Study of different fragmentation mode during the emission of multi charged projectile fragments reflects that the production of single projectile fragment with charge Z>2 is the most probable mode of multi-charge projectile fragments emission. This probability increases with the increase of projectile mass.

# Introduction

Heavy ion collisions provide us a unique opportunity to study the extended state of matter under extreme **conditions in terms of density** and temperature reachable otherwise in the hot early universe. At high densities and temperatures, a phase transition of hadronic **matter to quark–gluon plasma (QGP) has been tested in the experiments of RHIC [1-4] and LHC [5-7]. The subsequent investigation of the QGP is the main objective of heavy ion experiments at ultra-relativistic energies [8].** Apart from investigating the presence of QGP signal in heavy ion collisions, studies of nuclear fragmentation is also important. In comparison to the study of multiparticle production, the study of nuclear fragmentation received rather limited attention. **In order to have detail knowledge of high energy collisions investigations on nuclear fragmentation is also necessary**. Nuclear fragmentation has been considered to be one of the most important aspects of heavy-ion collisions, since it has been speculated that the decay of a highly excited nuclear system might carry information about the equation of state and the liquid–gas phase transition of low-density nuclear matter [**9-10**]. In nuclear fragmentation, both the target and the projectile can go into disintegration yielding target and projectile fragments respectively [**11**]. The fragmentation of relativistic projectile nucleus is of great significance as it leads to the emission of fragments with a broad mass spectrum which extends from the lightest fragments, the nucleons to the fragments as heavy as the disintegrating projectiles [**12-13**]. This allows us to extract information on the fragmentation mechanism involved in such processes. The study of projectile fragmentation reflects some important features of heavy ion collisions and has encouraged the scientists to perform experiments with projectile fragmentation [**14-27**]. Depending on the collision geometry, nucleus-nucleus collisions can be divided into three categories namely central collisions, quasi-central collisions and peripheral collisions. **In the central collisions, the projectile and target overlap, with a number of projectile spectator equal to zero and without the emission of projectile fragments.** Central collisions create favourable environment to study the QGP phase transition. As the centrality of the collision is decreased projectile spectators begin to increase in number. In peripheral collision the projectile and the target nuclei are far apart, so a small momentum is transferred **to** the interacting nuclei and the impact parameter is nearly equal to the sum of the radii of the target and the projectile nuclei. **The study of** **peripheral collisions has been overshadowed by the study of central and near central collisions and they have attracted a limited interest of the physicists.** Peripheral collision is the best situation of studying the projectile fragmentation. A detailed study of the projectile fragment makes it **possible to shed light** in the search for complicated quasi-stationary states of the fragments. In the nuclear scale of distances and excitations they can possess properties which make them analogous to dilute quantum gases in atomic physics at ultra-cold temperatures. The existence of such systems can find some important applications for the problems of nuclear astrophysics. In this respect, the fragment jets are a microscopic model of stellar media [**28**].

There is no doubt that the experiments performed **at** RHIC and **at the** LHC have many advantages in the study of relativistic particle production [**29-32**], which is related to the formation of quark-gluon plasma (QGP). However, for the study of projectile fragmentation, the accelerator experiment with the fixed target has more advantages. Goal of this study is the exploration of projectile fragmentation in the peripheral nucleus-nucleus collisions using nuclear emulsion track detector. In this paper we present the experimental results on the multiplicity distribution of singly charged, doubly charged and multi charged projectile fragments and also of shower particles in peripheral collisions for $^{16}$O-emulsion, $^{22}$Ne-emulsion and $^{28}$Si-emulsion interactions at an incident momentum of $(4.1-4.5)$ AGeV/c. The study of different fragmentation mode of multi charged fragment emission has also been performed. Nuclear emulsion detectors offer a good angular resolution (~0.1 mrad) and high spatial resolution which makes it possible to detect all the projectile fragments of charge $Z \geq 1$ in the **full solid angle** . This detector is well suited to study the different fragmentation mode during multi-fragment emission process.

**Experimental Details**

In order to collect **the data used for the** analysis **presented in this paper** NIKFI-BR2 emulsion pellicles of dimension 20cm$\times$10 cm $\times 600\,\mu$ m were irradiated by the $^{16}$O (at an incident momentum of 4.5 AGeV/c), $^{22}$Ne (at an incident momentum of 4.1 AGeV/c) and $^{28}$Si beam (at an incident momentum of 4.5 AGeV/c) accelerated from the Synchrophasotron at **the** Joint Institute of Nuclear Research (JINR), Dubna, Russia [**33**].

The emulsion pellicles **were exposed horizontally with respect to the direction of the incoming** projectile beams. When a projectile collides with the target present in nuclear emulsion an interaction or an event occurs. In emulsion detector scanning of the emulsion plate is performed to search for the primary interactions. We have searched for the primary interactions along the track of the incident beam using an objective of 100X magnification of a usual scanning microscope. We have applied the double scanning method along the track of the incident projectile beam fast in the forward direction (i.e. from the entry point of the beam into the emulsion plate till an interaction occurs) and slowly in the reverse direction [**33**]. Slow scanning in the backward direction ensures that the events chosen did not include interactions from the secondary tracks of the other interactions. Scanning in the forward and backward directions by two different observers increases the efficiency of selecting a primary event up to 99% [**33**].

**After a scanning phase** events were selected according to the criteria mentioned below:

   a) The incident beam track would have to lie within 3∘ from direction of the main beam in the pellicle. This criterion **ensures** that the real projectile beam has been selected for the analysis.

   b) Events showing interactions close to the emulsion surface and glass surface (interactions within 20 $\mu$m from the top and bottom surface of the pellicle) were excluded from **our consideration. Rejection** of such events reduces the losses of tracks and minimizes the uncertainties in the measurements of emission and azimuthal angles [**33**].

According to the criteria mentioned above, we have **selected** 2823 inelastic interactions for the $^{16}$O projectile, 4308 inelastic interactions of the $^{22}$Ne projectile and 1310 inelastic interactions of the $^{28}$Si projectile.

According to the Powell [**34**], in nuclear emulsion track detector, particles emitted and produced from an interaction can be classified into four categories, namely the projectile fragments, shower particles, the grey particles and the black particles. **Details about these categories of particles have been reported in the following**.

**Projectile Fragments**: The projectile fragments are the spectator parts of the incident projectile nucleus that do not directly participate in **the** interaction. They are emitted within a forward cone characterized by the semi vertex angle $\theta_f < 3^\circ$. The semi vertex angle $\theta_f$ can be defined as $\theta_f = \frac{P_{Fermi}}{P_{inc}}$ [**35**] where $P_{inc}$ is the incident projectile momentum per nucleon measured in GeV/c and $P_{Fermi}$ is the Fermi momentum of the nucleons of the projectile in GeV/c. **The value of** $P_{Fermi}$ **can be calculated on the basis of Fermi gas model of the nucleus [36].** Having almost the same energy or momentum per nucleon as the incident projectile, these fragments exhibit uniform ionization over a long range and suffer negligible scattering. The main sources of the projectile fragments are nucleons and nucleon clusters formed in the nuclear collisions. Projectile fragments can be further classified according to their charges as singly charged (Z=1), doubly charged (Z=2) and multi-charged (Z >2) projectile fragments.

(i) Singly charged projectile fragments have ionisation I less or equal to $1.4I_0$. $I_0$ is the minimum ionization of a singly charged particle. Multiplicity of singly charged fragments is denoted by $N_P$.

(ii) Doubly charged projectile fragments are denoted by $N_\alpha$. They have ionisation I $\cong 4I_0$ and no change in their ionization can be noticed when the tracks are followed up to a distance of approximately 2 cm from the vertex of interaction.

(iii) Multi-charged projectile fragments are denoted by $N_F$. They have ionisation I > $6I_0$ and no change in their ionization can be noted when their tracks are followed up to a distance of approximately 1 cm from the vertex of interaction. Charge of these fragments can be measured by the delta ray counting method [**37-38**]**.**

**Shower particles**: The tracks of particles having ionization I less or equal to $1.4I_0$ are called shower tracks. The shower particles are mostly pions (about more than 90%) with a small **percentage** of kaons and hyperons (less than 10%). These shower particles are produced in a forward cone. The velocities of these particles are greater than $0.7c$ where $c$ is the **speed** of light in free space. Because of such a high velocity, these particles are not generally confined within the emulsion pellicle. Energies of these shower particles lie in the GeV range. Average multiplicity of shower particles can be denoted by $\langle N_s \rangle$.

**Grey particles**: Grey particles are mainly fast target recoil protons with energies up to 400 MeV. They have ionization $1.4\ I_0 \leq I < 10\ I_0$. Ranges of these particles are greater than 3 mm in the emulsion medium. **They are characterized by velocity lying between $0.3\,c$ and** $0.7\,c$. Average multiplicity of grey particles is denoted by $\langle N_g \rangle$.

**Black particles**: In this category **both singly and multiply charged fragments are included**. They are fragments of various elements like carbon, lithium, beryllium etc with ionization greater or equal to $10 I_0$. **Black particles have the maximum ionizing power but they are also less energetic and consequently short ranged particles. In the emulsion medium, ranges of black particles are less than 3 mm and the velocities of these particles are less than $0.3\,c$.** In emulsion experiments, it is very difficult to measure the charges of the target **fragments and hence it is not possible to identify the exact nucleus. Average multiplicity of** black particles is denoted by $\langle N_b \rangle$. Total number of black and grey tracks in an event is known as heavy tracks and is denoted by $N_h$.

After finding the primary inelastic interactions induced by the incoming projectiles, the number of black, grey, shower tracks and the number of projectile fragments with charge Z=1,Z=2 and Z>2 were counted for all the events of each interaction using an oil immersion objective of 750X magnification of a special microscope KSM (Kernspurmessmikroskop) made by Karl Zeiss Jena [**33**].

**Results and Analysis**

In this paper we have designated the peripheral events as the events having the total number of black and grey tracks to be zero that is $N_h=0$. Applying this criteria we have selected peripheral events in $^{16}$O-emulsion, $^{22}$Ne-emulsion and $^{28}$Si-emulsion interactions at an incident momentum of (4.1−4.5) AGeV/c. Along with the number of peripheral events, we have also calculated the percentage of peripheral events (POE) in each interaction and presented the values in table 1. From the table it is seen that percentage of peripheral events (POE) increases with the increase of mass number $A_P$ of the projectile beam. Figure 1 shows the variation of percentage of peripheral events (POE) with the mass number of the projectile beam for $^{16}$O-emulsion, $^{22}$Ne-emulsion and $^{28}$Si-emulsion interactions at an incident momentum of (4.1−4.5) AGeV/c. **Uncertainties** shown in the figure are statistical **uncertainties** only. We have tried to fit the variation of percentage of peripheral events (POE) with the mass number of the projectile beam $A_P$ with a function of the form $POE = aA_P{}^b$. The values of the fitting parameters a and b have been calculated on the basis of the $\chi^2$ minimization method. Calculated values of a and b have been found to be $a = 6.74 \pm 0.06$ and $b = 0.198 \pm 0.003$ with $\chi^2$ per degrees of freedom values 0.98.

The average multiplicity of singly charged ($N_P$) doubly charged ($N_\alpha$) and multi charged projectile fragments ($N_F$) along with the produced shower particles in the peripheral events have been presented in table 2 for the three interactions. From the table it may be noted that within the experimental uncertainty average multiplicity of singly charged ($N_P$) and doubly charged ($N_\alpha$) fragments show an increasing trend with the increase of projectile mass. Moreover from the table it may be noted that for each projectile the values of average multiplicity of projectile fragments for Z=1 are higher than those for Z=2 and Z>2.

From the table it may also be noted that in case of shower particles the average multiplicity increases smoothly with the increase of projectile mass. We have studied the variation of the average multiplicity of shower particles $\langle N_s \rangle$ with the mass number of the projectile beam $A_p$. The variation of average shower particle multiplicity with mass number of the projectile beam has been presented in figure 2. **Uncertainties** shown in figure 2 are of statistical origin. **The uncertainties of average multiplicity of each interaction have been calculated from the standard deviation of the event sample.** The variation of $\langle N_s \rangle$ with $A_p$ has been parametrized by a relation $\langle N_s \rangle = aA_p^b$. The values of the fitting parameters have been evaluated on the basis of the $\chi^2$ minimization method. We have also calculated the values of the non-linear regression coefficient $\left(R^2\right)$ for the fits in order to quantify the goodness of the fit. The values of the fitting parameters, $\chi^2$ per degrees of freedom ($\chi^2$ /DOF) value of the fit and the values of non-linear regression coefficient $\left(R^2\right)$ of the fit are **reported** in table 3. The exponent b characterises the increment of average shower particle multiplicity with the mass of the projectile beam. In a very recent paper [**39**] we have shown that in central events of $^{16}$O-AgBr, $^{22}$Ne-AgBr and $^{28}$Si-AgBr interactions at an incident momentum of $(4.1-4.5)$ AGeV/c, the increment of $\langle N_s \rangle$ with $A_P$ can be parametrized as $\langle N_s \rangle = (3.34 \pm 0.11)A_P^{(0.75\pm0.08)}$. On the contrary for peripheral events average shower particle multiplicity increases with $A_P$ as $\langle N_s \rangle = (1.23 \pm 0.09)A_P^{(0.24\pm0.07)}$. Thus for peripheral events increment of $\langle N_s \rangle$ with $A_P$ is rather slow in comparison to central events.

Average multiplicity of multi-charged projectile fragments (Z>2) are found to increase with the increase of projectile mass as evident from table 2. **We have investigated the variation of the average multiplicity of multi-charged projectile fragments with the mass number of the projectile beam $A_p$ in figure 3**. **Uncertainties** associated with the experimental points are the statistical **uncertainties** only. The variation has been parametrized by a relation $\langle N_F \rangle = aA_p^b$. The method of the $\chi^2$ minimization has been applied in this case also to extract the values of the fitting parameters a and b. The values of the non-linear regression coefficient $\left(R^2\right)$ for the fits have also been calculated. The values of the fitting parameters, $\chi^2$ per degrees of freedom ($\chi^2$ /DOF) value of the fit and the values of non-linear regression coefficient $\left(R^2\right)$ of the fit are given in table 3. **Taking into the account the values reported in table 3, the average multiplicity of multi-charged fragments can be expressed as function of the projectile mass $A_P$ according to the relation $\langle N_F \rangle = (0.065 \pm 0.005)A_P^{(0.76\pm0.07)}$.**

Multiplicity distribution of projectile fragments having charge Z=1 and Z=2 along with the multiplicity distribution of shower particles for the peripheral events have been presented in fig 4(a)-4(c) for $^{16}$O-emulsion interactions, in fig 5(a)-5(c) for $^{22}$Ne-emulsion interactions and in fig 6(a)-6(c) for $^{28}$Si-emulsion interactions. In the multiplicity distribution plots the probability $P_n$ has been calculated according to the relation $P_n = \dfrac{\text{Number of events having n number of tracks}}{\text{Total number of events}}$. From the multiplicity distribution of the projectile fragments it may also been pointed out that the probability of producing singly charged and doubly charged projectile fragments decreases as the number of these fragments increases. In table 4 we have presented the values of the maximum number of projectile fragments with charge Z=1 and Z=2 emitted in a single event for each of the projectiles. Table 4 reflects that maximum number of singly charged (Z=1) and doubly charged (Z=2) projectile fragments emitted in single event is highest for Si projectile.

**From the figure 4(a)-4(c) it may be noted that the maximum value of the probability distribution ($P_n^{max}$) of singly charged fragments decreases** with the increase of projectile mass. However for doubly charged fragments no such observation can be made as evident from figure 5(a)-5(c) where **maximum value of the probability distribution** occurs for $^{22}$Ne-emulsion interactions. **No systematic dependence of the maximum value of the probability distribution of shower particles on the mass number of the projectile beam can be noted from figure 6(a)-6(c).** The multiplicity distributions of multi-charged fragments for the three interactions have been shown in fig7. From the figure it is clear that **maximum value of the probability distribution** for multi-charged fragments increases with the increase of projectile mass.

In order to study the multiplicity distribution of the singly charged, doubly charged, multi charged projectile fragments and the shower particles in peripheral events of the three interactions in more details we have calculated the dispersion of the multiplicity distribution following the relation $D = \sqrt{\langle N^2 \rangle - \langle N \rangle^2}$. The width of the multiplicity distribution is characterised by the parameter D. The calculated values of the dispersion D have been presented in table 5 for singly charged , doubly charged , multi charged projectile fragments and shower particles in peripheral events in case of $^{16}$O-emulsion,$^{22}$Ne-emulsion and $^{28}$Si-emulsion interactions at an incident momentum of $(4.1-4.5)$ AGeV/c. From the table it may be seen that with the increase of projectile mass the values of the dispersion increases for the singly charged, doubly charged projectile fragments and for the shower particles.

However for multi charged fragments dispersion D decreases with the increase of projectile mass.

In order to understand the dynamics of projectile fragmentation we have also studied the fragmentation mode of multi-fragment emission in the peripheral events of $^{16}$O-emulsion, $^{22}$Ne-emulsion and $^{28}$Si-emulsion interactions at an incident momentum of (4.1−4.5) AGeV/c. In table 6 we have presented the values of the probability of producing one projectile fragments with charge Z>2 and two projectile fragments with charge Z>2 along with the probability of producing projectile fragments with charge Z>2 with and without the emission of any α particles. **The results reported in the table suggest that the probability of producing one projectile or two projectiles with Z>2 and of producing fragments with Z>2 with alpha particle emission increases when the projectile mass increases.**

While the probability of emitting projectile fragments with charge Z>2 without the emission of any α particles do not show any systematic variation with the mass number of the projectile beam. Moreover it is evident from the table that the probability of emitting more than one fragment with charge Z>2 is significantly smaller than that of emitting one projectile fragment with charge Z>2. This **interpretation** has also been supported in [**40**]. From table 6 it may be noticed that for $^{22}$Ne-emulsion and $^{28}$Si-emulsion interactions the emission of projectile fragments with charge Z>2 accompanied by the emission of α particles is more probable than that of without the emission of α particles. On the contrary for $^{16}$O-emulsion interactions the probability of emission of projectile fragments with charges Z>2 without the emission of α particles is higher in comparison to the emission of Z>2 charged fragments accompanied by the emission of α particles.

### Discussion on systematic Uncertainties

Earlier we have mentioned that the uncertainties presented in all the tables and the plots are statistical uncertainties only. Apart from the statistical uncertainties, it is necessary to study the influence of the systematic uncertainties on our analysis. The systematic uncertainties may arise during the scanning of the emulsion plates. It has been mentioned earlier that in order to find all the primary interactions "along the track" scanning technique was applied. "Along the track" scanning method gives reliable event samples because of its high detection efficiency [**41-43**]. It has already been discussed that the scanning was done fast in the forward direction and slowly in the backward direction. Two independent observers scanned each plate so that the biases in detection, counting and measurements can be minimized. This process helps us to obtain a scanning efficiency more than 99%. Therefore, the contributions to the systematic uncertainties arising from the scanning procedure are less than 1%. Systematic uncertainties may also be introduced due to the presence of background tracks in nuclear emulsion detector. The background tracks can arise due to the decay of radioactive contamination, such as thorium present both in emulsion and in their glass backings [**44**]. Some of such tracks may also be caused by the cosmic radiations during the time of exposure. But to confuse a background track with the tracks coming from an event, the track has to originate from the interaction vertex – the chance of which is only accidental. Moreover, the volume occupied by all interaction vertices in an emulsion plate is negligibly small with respect to the total volume of the plate. Keeping these points in mind, the chance of a mix-up between tracks originating from an event, with the tracks caused by background radiations are very small. To the readers the issue of background contamination due to production of electron pair tracks may be an important issue to be addressed. However, the electron pair tracks produced through $\gamma$ - conversion do not emanate from the interaction vertex, but are produced at a distance from the vertex after traveling through certain radiation lengths. In order to eliminate all the possible backgrounds due to $\gamma$ overlap (where a $\gamma$ from a $\pi^0$ decay converts into $e^+e^-$ pair) close to shower tracks near vertex, special care was taken to exclude such $e^+e^-$ pairs from the primary shower tracks while performing angular measurements. Usually all shower tracks in the forward direction were followed more than 100–200 μm from the interaction vertex for angular measurement. The tracks due to $e^+e^-$ pair can be easily recognized from the grain density measurement, which is initially much larger than the grain density of a single charged pions or proton track. It may also be mentioned that the tracks of an electron and positron when followed downstream in nuclear emulsion showed considerable amount of Coulomb scattering as compared to the energetic charged pions. Such $e^+e^-$ pairs were eliminated from the data. Apart from the background contamination, systematic uncertainties may also arise in the measurement of polar angles due to fading of tracks. The rate of physical fading is very little at normal temperature (0∘C–25∘C). Reported values indicate that it is less than $7 \times 10^{-4}$ reductions per day [**45**]. To prevent thermal fading, the exposed emulsions are kept at very low temperature. Thus, the uncertainties arising from the fading of tracks are found to be relatively small. So from our discussion on systematic uncertainties of nuclear emulsion track detector, it is clear that the systematic uncertainties present in our analysis are of the order of 1%.

**Conclusions and Outlook**

**We have presented a study of multiplicity distribution of singly charged, doubly charged and multi charged projectile fragments and also of shower particles in peripheral events of** $^{16}$O-emulsion, $^{22}$Ne-emulsion and $^{28}$Si-emulsion interactions at an incident momentum of $(4.1-4.5)$ AGeV/c. Events with complete absence of target fragmentation have been selected as peripheral events. Study of different fragmentation mode of multi-charged projectile fragments emission has also been presented. **Qualitative information about the fragmentation of relativistic nuclei in peripheral collisions have been extracted and reported.** The emulsion technique allows one to observe these systems to the smallest details and gives the possibility of studying them experimentally. **The observed results of this study of multiplicity distribution of projectile fragments and shower particles in peripheral events of nucleus-nucleus collisions can be viewed as an experimental fact.**

The significant conclusions of this **analysis can be summarized as follows**.

1. Percentage of peripheral events (POE) selected according to the criteria $N_h = 0$ increases with the increase of projectile mass $A_P$ according to the relation $POE = (6.74 \pm 0.06)A_P^{(0.198 \pm 0.003)}$.

2. Average multiplicities of projectile fragments **with** charge Z=1 and Z=2 show a slight increase with the increase of projectile mass within the experimental **uncertainties**. However, average multiplicity of multi-charged projectile fragments increases with $A_P$ according to the relation $\langle N_F \rangle = (0.065 \pm 0.005)A_P^{(0.76 \pm 0.07)}$. In case of shower particle average multiplicity increases smoothly with the mass number of the projectile beam $A_P$ according to the relation $\langle N_s \rangle = (1.23 \pm 0.09)A_P^{(0.24 \pm 0.07)}$.

3. Probability of producing singly charged, doubly charged and multi-charged projectile fragments decreases as the number of these fragments increases. Maximum number of singly charged and doubly charged projectile fragments emitted in a single event is highest for the heaviest projectile among the three projectiles.

4. Maximum value of the probability distribution for the singly charged fragments decreases with the increase of projectile mass while that increases with the increase of projectile mass for multi-charged fragments.

5. The dispersion of multiplicity distribution for singly charged, doubly charged projectile fragments and also for the shower particles increase with the increase of projectile mass. Dispersion values of the multiplicity distribution for the multi-charged fragments decreases with the increase of projectile mass.

6. Production of single projectile fragments with charge Z>2 is the most probable mode of multi-charge projectile fragments emission. Probability of emission of two projectile fragments with charge Z>2 is significantly small. Both the probabilities increase with the increase of projectile mass. Emission of projectile fragments **with** charge Z>2 accompanied by the emission of α particle is more probable than that of without the emission of α particle for $^{22}$Ne-emulsion and $^{28}$Si-emulsion interactions. Probability of emission of projectile fragments **with** charge Z>2 accompanied by the emission of α particle increases with the increase of projectile mass. Probability of emission of projectile fragments **with** charge Z>2 without the emission of α particle is independent of the mass of the projectile beam.

**There are many papers available in the literature where a detailed study of projectile fragmentation has been carried out but most of the papers are not related to peripheral collisions, so our paper represents a step forward in such a field.**

**Acknowledgement**

The authors are grateful to Prof. Pavel Zarubin, JINR, Dubna, Russia for providing them the required emulsion data. Dr. Bhattacharyya also acknowledges Prof. Dipak Ghosh, Department of Physics, Jadavpur University and Prof. Argha Deb Department of Physics, Jadavpur University, for their inspiration in the preparation of this manuscript.

**Table 1**

| Interactions | Total inelastic interactions | Number of peripheral events | Percentage of peripheral events |
|---|---|---|---|
| $^{16}$O-emulsion (4.5 AGeV/c) | 2823 | 330 | $(11.68\pm0.10)\%$ |
| $^{22}$Ne-emulsion (4.1 AGeV/c) | 4308 | 537 | $(12.46\pm0.14)\%$ |
| $^{28}$Si-emulsion (4.5 AGeV/c) | 1310 | 171 | $(13.05\pm0.16)\%$ |

Table 1 represents the total inelastic interactions, number of peripheral events and Percentage of peripheral events for $^{16}$O-emulsion, $^{22}$Ne-emulsion and $^{28}$Si-emulsion interactions at an incident momentum of (4.1−4.5) AGeV/c.

**Table 2**

| Interactions | Average Multiplicity of singly charged fragments $\langle N_P \rangle$ | Average Multiplicity of Doubly charged fragments $\langle N_\alpha \rangle$ | Average Multiplicity of Multi-Charged charged fragments $\langle N_F \rangle$ | Average multiplicity of shower particles $\langle N_s \rangle$ |
|---|---|---|---|---|
| $^{16}$O-emulsion (4.5 AGeV/c) | 1.57$\pm$0.03 | 1.22$\pm$0.02 | 0.52$\pm$0.05 | 2.49$\pm$0.11 |
| $^{22}$Ne-emulsion (4.1AGeV/c) | 1.49$\pm$0.14 | 1.09$\pm$0.21 | 0.76$\pm$0.06 | 2.60$\pm$0.15 |
| $^{28}$Si-emulsion (4.5 AGeV/c) | 2.49$\pm$0.15 | 1.44$\pm$0.31 | 0.82$\pm$0.09 | 2.86$\pm$0.19 |

Table 2 represents the values of average Multiplicity of singly charged fragments $\langle N_P \rangle$, average Multiplicity of Doubly charged fragments $\langle N_\alpha \rangle$, average Multiplicity of Multi-Charged charged fragments $\langle N_F \rangle$ and average multiplicity of shower particles $\langle N_s \rangle$ for $^{16}$O-emulsion,$^{22}$Ne-emulsion and $^{28}$Si-emulsion interactions at an incident momentum of (4.1−4.5) AGeV/c for the peripheral events (events without target fragmentation , $N_h$ =0).

**Table 3**

| Particles | $a$ | $b$ | $\chi^2 / DOF$ | $R^2$ |
|---|---|---|---|---|
| Shower particles | 1.23±0.09 | 0.24±0.07 | 0.97 | 0.99 |
| Multi-Charged Fragments(Z>2) | 0.065±0.005 | 0.76±0.07 | 0.94 | 0.94 |

Table 3 represents the values of the fitting parameters of the plot of average multiplicity with the mass number of the projectile $\langle N_s \rangle = a A_P^b$ in case of the shower particle multiplicity and values of the fitting parameters of the plot of average multiplicity with the mass number of the projectile beam for the multi-charged fragments $\langle N_F \rangle = a A_P^b$ , the $\chi^2$ per degrees of freedom ( $\chi^2$ /DOF) value of both the fit and the non-linear regression coefficient $R^2$ for both the fits in case of $^{16}$O-emulsion,$^{22}$Ne-emulsion and $^{28}$Si-emulsion interactions at an incident momentum of (4.1−4.5) AGeV/c for the peripheral events (events without target fragmentation , $N_h$ =0) in our analysis.

**Table 4**

| Interactions | Z=1 (maximum value in a event) | Z=2 (maximum value in a event) |
|---|---|---|
| $^{16}$O-emulsion (4.5 AGeV/c) | 6 | 4 |
| $^{22}$Ne-emulsion (4.1AGeV/c) | 8 | 5 |
| $^{28}$Si-emulsion (4.5 AGeV/c) | 9 | 6 |

Table 4 represents the values of the maximum number of projectile fragments of both Z=1 and Z=2 emitted in a single peripheral event (event without target fragmentation, $N_h$=0) for $^{16}$O-emulsion,$^{22}$Ne-emulsion and $^{28}$Si-emulsion interactions at an incident momentum of (4.1−4.5) AGeV/c.

**Table 5**

| Interactions | Dispersion Values  $D= \sqrt{\langle n^2 \rangle - \langle n \rangle^2}$ | | | |
|---|---|---|---|---|
| | Singly Charged Projectile Fragments | doubly Charged Projectile Fragments | Multi-Charged Projectile Fragments | Shower Particles |
| $^{16}$O-emulsion | 1.34±0.02 | 1.13±0.01 | 0.499±0.010 | 2.20±0.08 |
| $^{22}$Ne-emulsion | 1.47±0.05 | 1.14±0.03 | 0.476±0.012 | 2.40±0.11 |
| $^{28}$Si-emulsion | 1.82±0.11 | 1.37±0.09 | 0.422±0.015 | 2.60±0.14 |

Table 5 represents the dispersion values of the multiplicity distribution for the singly charged, doubly charged, multi-charged projectile fragments and shower particles in $^{16}$O-emulsion, $^{22}$Ne-emulsion and $^{28}$Si-emulsion interactions at an incident momentum of (4.1—4.5) AGeV/c for the peripheral events (events without target fragmentation , $N_h$ =0).

**Table 6**

| Fragmentation Mode | $^{16}$O- emulsion | $^{22}$Ne- emulsion | $^{28}$Si-emulsion |
|---|---|---|---|
| One Projectile Fragments with Z>2 | 51.81% | 72.47% | 76.60% |
| Two Projectile Fragments with Z>2 | 0% | 1.12% | 2.33% |
| Projectile Fragments with Z>2 and α particles | 22.72% | 41.94% | 51.46% |
| Projectile Fragments with Z>2 and no α particles | 29.09% | 33.02% | 28.07% |

Table 6 represents the percentage of different fragmentation mode of multi-charged projectile fragments in peripheral collisions of $^{16}$O-emulsion, $^{22}$Ne-emulsion and $^{28}$Si-emulsion interactions at an incident momentum of (4.1−4.5) AGeV/c.

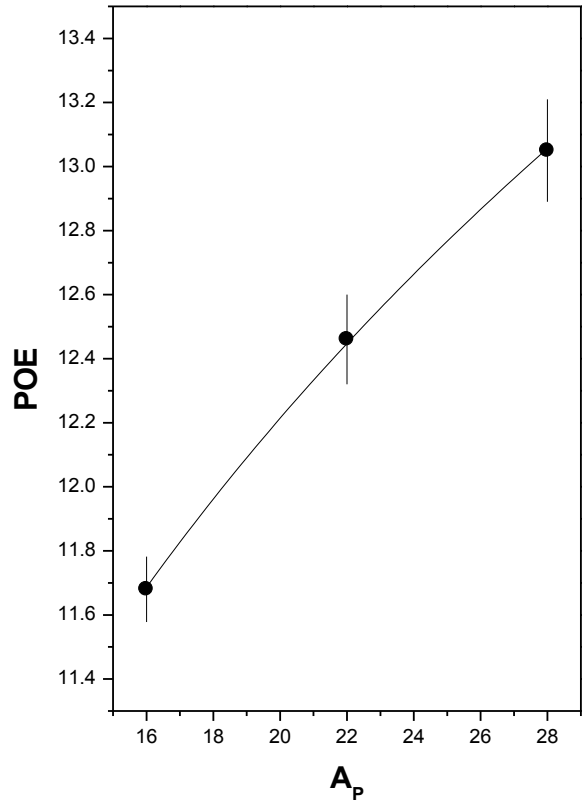

Fig1 Variation of the percentage of peripheral events (events without target fragmentation, $N_h$ =0) with the mass number of the projectile beam for $^{16}$O-emulsion, $^{22}$Ne-emulsion and $^{28}$Si-emulsion interactions at an incident momentum of $(4.1-4.5)$ AGeV/c. The solid line represents the non linear fit $POE = aA_P{}^b$.

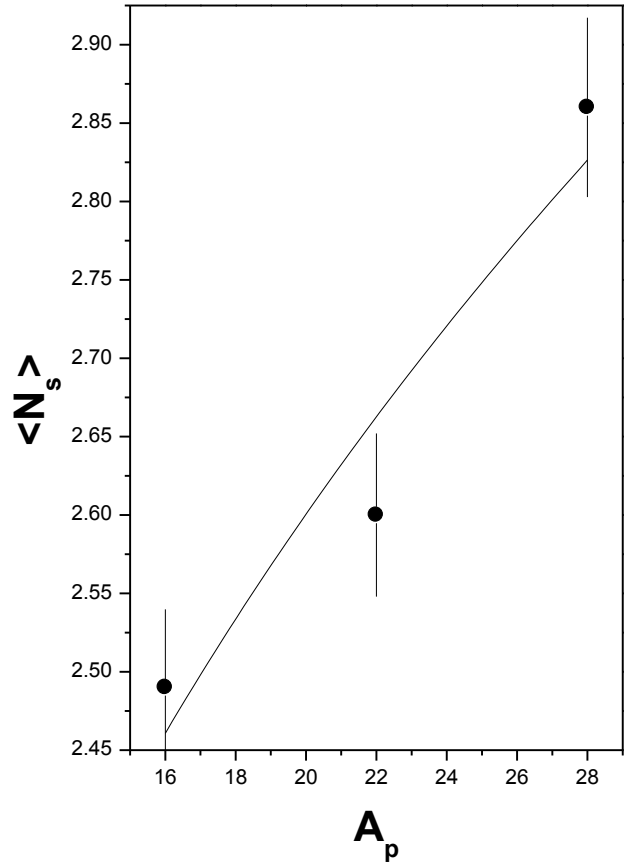

Figure 2 The variation of average multiplicity of shower particles with the mass number of the projectile beam for $^{16}$O-emulsion, $^{22}$Ne-emulsion and $^{28}$Si-emulsion interactions at an incident momentum of (4.1−4.5) AGeV/c for the peripheral events (events without target fragmentation, $N_h$ =0). The solid line represents the non linear fit $\langle N_s \rangle = aA_p{}^b$.

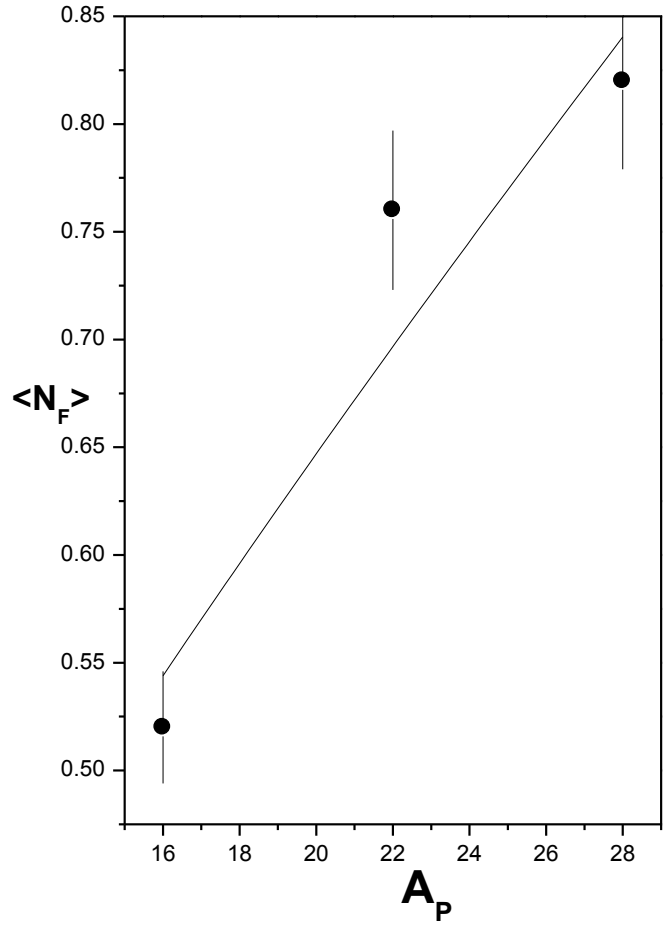

Figure 3 The variation of average multiplicity of multi-charged fragments (Z>2) with the mass number of the projectile beam $A_P$ for $^{16}$O-emulsion,$^{22}$Ne-emulsion and $^{28}$Si-emulsion interactions at an incident momentum of (4.1−4.5) AGeV/c for the peripheral events (events without target fragmentation, $N_h$ =0). The solid line represents the non linear fit $\langle N_F \rangle = aA_P{}^b$.

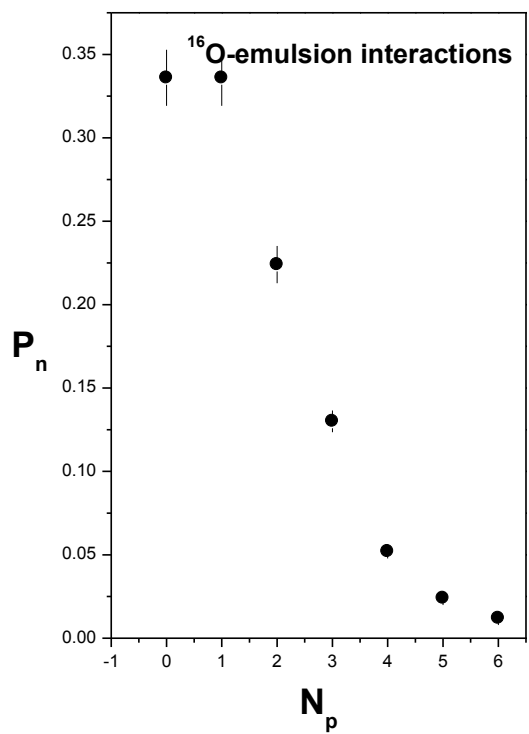

Figure 4(a) Multiplicity distribution of projectile fragments with Z=1 for $^{16}$O-emulsion interactions at an incident momentum of 4.5 AGeV/c for the peripheral events (events without target fragmentation, $N_h$ =0).

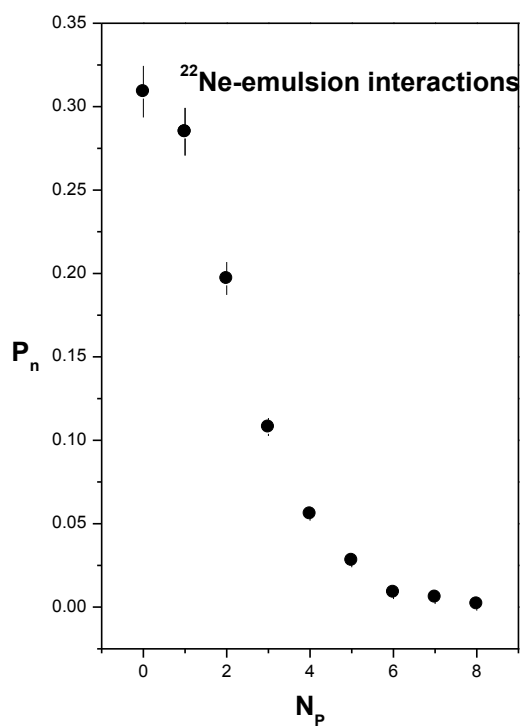

Figure 4(b) Multiplicity distribution of projectile fragments with Z=1 for $^{22}$Ne-emulsion interactions at an incident momentum of 4.1 AGeV/c for the peripheral events (events without target fragmentation, $N_h$ =0).

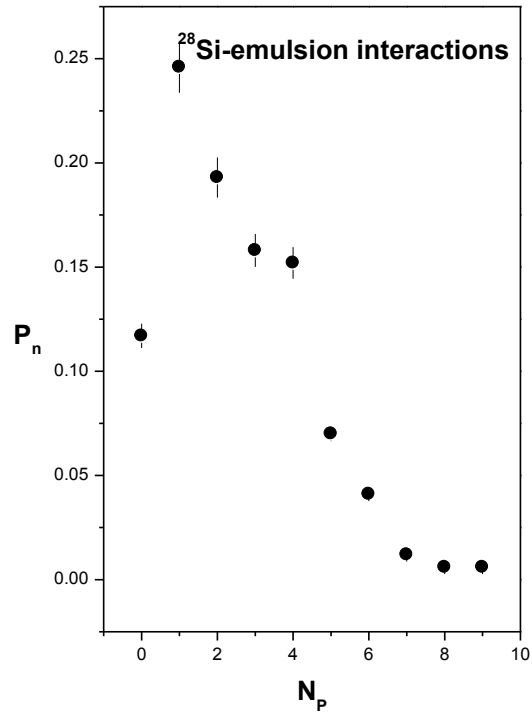

Figure 4(c) Multiplicity distribution of projectile fragments with Z=1 for $^{28}$Si-emulsion interactions at an incident momentum of 4.5 AGeV/c for the peripheral events (events without target fragmentation, $N_h$ =0).

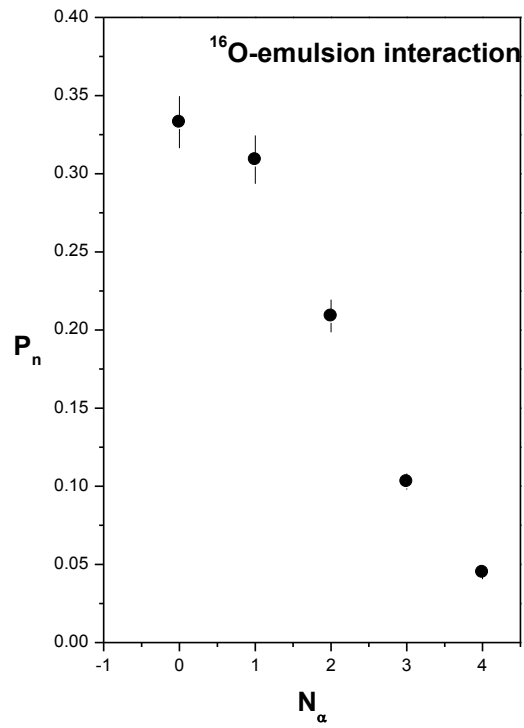

Figure 5(a) Multiplicity distribution of projectile fragments with Z=2 for $^{16}$O-emulsion interactions at an incident momentum of 4.5 AGeV/c for the peripheral events (events without target fragmentation, $N_h$ =0).

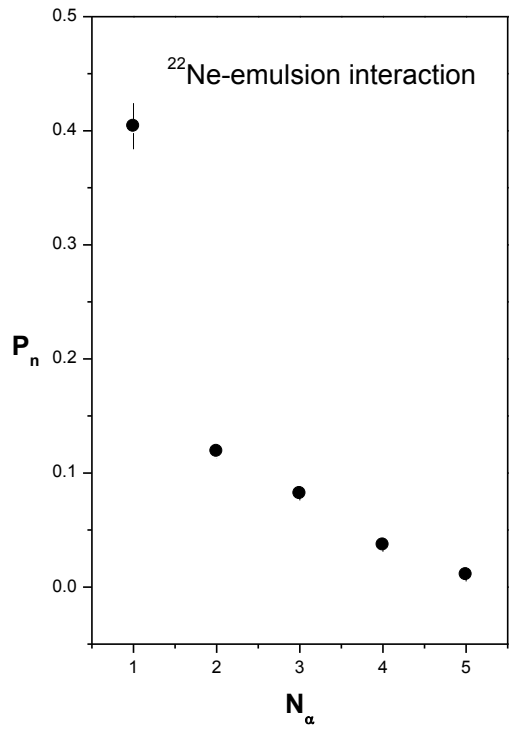

Figure 5(b) Multiplicity distribution of projectile fragments with Z=2 for $^{22}$Ne-emulsion interactions at an incident momentum of 4.1 AGeV/c for the peripheral events (events without target fragmentation, $N_h$ =0).

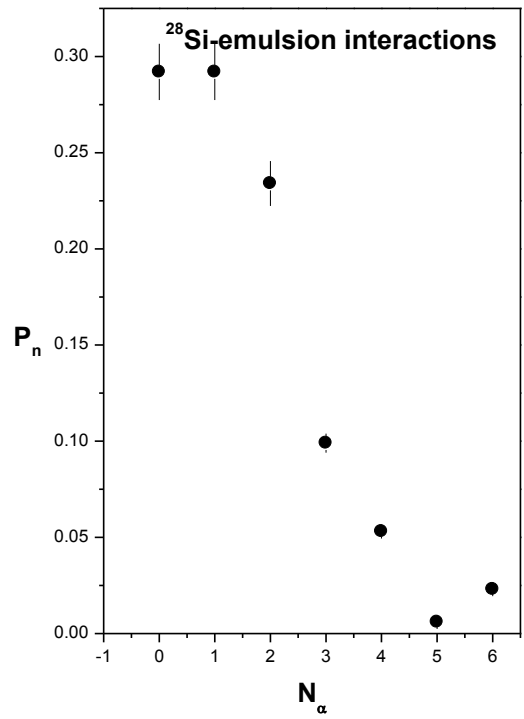

Figure 5(c) Multiplicity distribution of projectile fragments with Z=2 for $^{28}$Si-emulsion interactions at an incident momentum of 4.5 AGeV/c for the peripheral events (events without target fragmentation, $N_h$ =0).

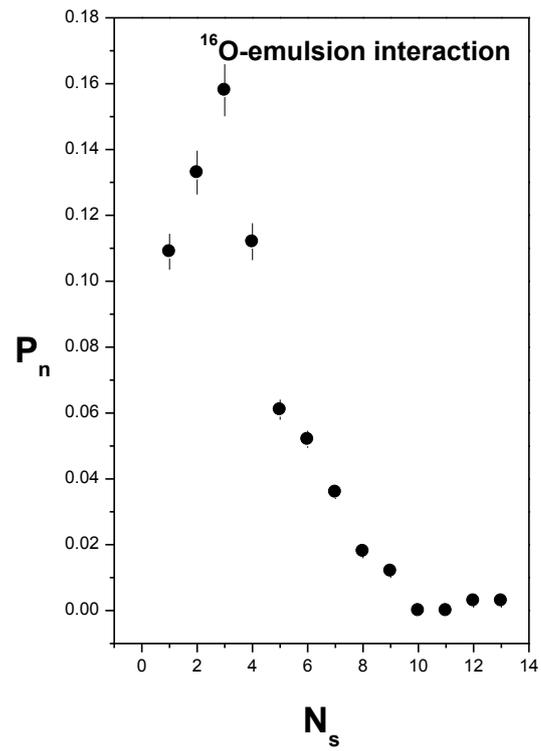

Figure 6(a) Multiplicity distribution of shower particles for $^{16}$O-emulsion interactions at an incident momentum of 4.5 AGeV/c for the peripheral events (events without target fragmentation, $N_h = 0$).

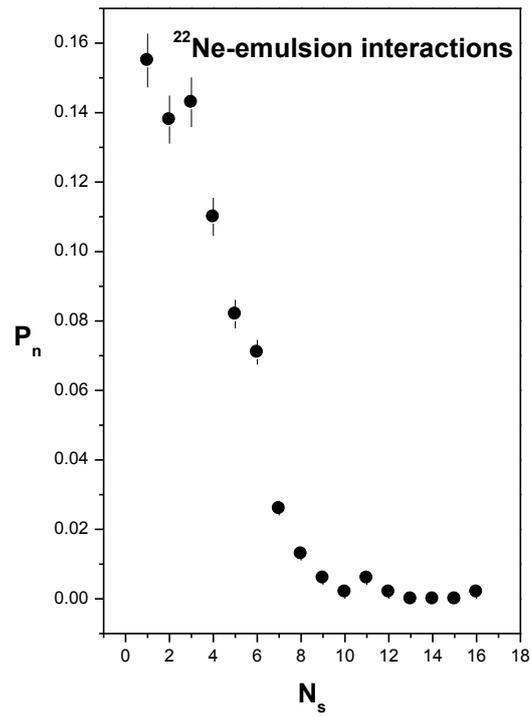

Figure 6(b) Multiplicity distribution of shower particles for $^{22}$Ne-emulsion interactions at an incident momentum of 4.1 AGeV/c for the peripheral events (events without target fragmentation, $N_h$ =0).

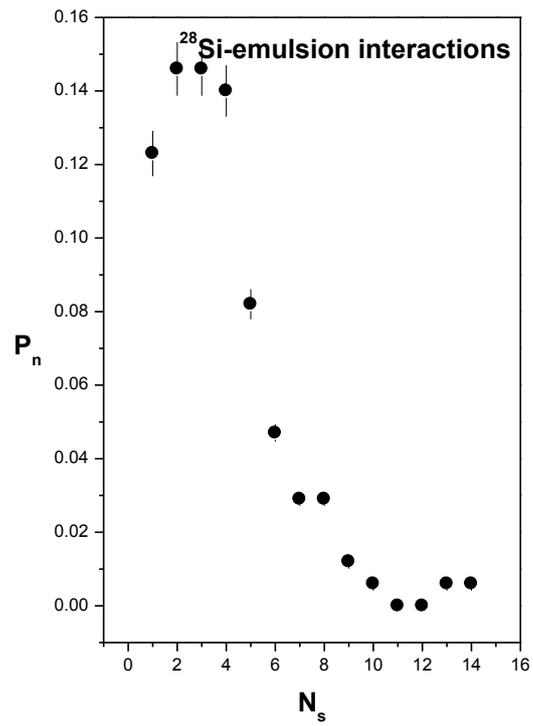

Figure 6(c) Multiplicity distribution of shower particles for $^{28}$Si-emulsion interactions at an incident momentum of 4.5 AGeV/c for the peripheral events (events without target fragmentation, $N_h = 0$).

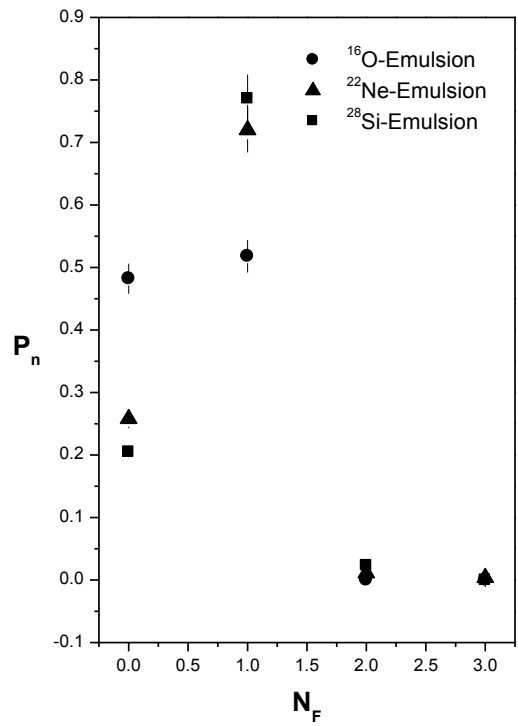

Figure 7   Multiplicity distribution of multi-charged   projectile fragments   of $^{16}$O-emulsion,$^{22}$Ne-emulsion and $^{28}$Si-emulsion interactions at an incident momentum of (4.1—4.5) AGeV/c for the  peripheral events (events without target fragmentation , $N_h$ =0).